\documentclass[a4paper]{article}
\usepackage{ISCSLP2026}
\usepackage{iftex}
\ifPDFTeX\else
  \usepackage{fontspec}
\fi
\usepackage{ifthen}
\newboolean{blind}
\setboolean{blind}{false}

\usepackage{cite}
\usepackage{amsmath,amssymb,amsfonts}
\usepackage{graphicx}
\usepackage{textcomp}
\usepackage{xcolor}
\usepackage{booktabs}
\usepackage{bm}
\usepackage{multirow}
\usepackage{tikz}
\usetikzlibrary{positioning, arrows.meta, fit, backgrounds, calc, decorations.pathreplacing}
\usepackage{pgfplots}
\pgfplotsset{compat=1.18}
\usepackage{comment}

\newcommand{\method}{LatentASR}
\newcommand{\vdelta}{\boldsymbol{\delta}}

\title{Listen, Think, Transcribe: Continuous Latent Test-Time Scaling for ASR}

\name{
	\ifthenelse{\boolean{blind}}{Anonymous to ISCSLP}
	{Ho Lam Chung$^{1,2}$, Yiming Chen$^{3}$, Dau-Cheng Lyu$^{2}$, Hsiao-Tsung Hung$^{2}$, Hung-yi Lee$^{1}$}
}

\address{
  \ifthenelse{\boolean{blind}}{Anonymous to ISCSLP}
  {
  	$^1$National Taiwan University, Taiwan \quad
  	$^2$ASUS, Taiwan \quad
  	$^3$National University of Singapore, Singapore
  }
}

\email{
	\ifthenelse{\boolean{blind}}{Anonymous to ISCSLP}
	{holam.chung@protonmail.com, yiming.chen@u.nus.edu, ricer\_lu@asus.com,
	 Alexht\_Hung@asus.com, tlkagkb93901106@gmail.com}
}

\begin{document}

\maketitle

\begin{abstract}
End-to-end ASR models transcribe in a single pass, leaving no room for the decoder to revisit hard inputs. We propose \emph{LatentASR}, a parameter-efficient method that adds continuous latent test-time scaling to a frozen ASR backbone. Two small trainable modules drive it: a Latent Adapter that iteratively refines a few latent prefix positions through bounded, stabilized updates, and a Value Head that predicts whether extra computation will help and halts the loop early. The Qwen3-ASR-0.6B backbone stays fully frozen, and we train only ${\sim}4$M extra parameters. We activate this loop with a deliberately small, diverse 500-utterance training set. Under this minimal-data regime, standard adaptation methods all regress: full fine-tuning, LoRA, and prompt tuning each increase WER. \method{} is the only tested method that reduces WER on both clean benchmarks (FLEURS $-2.54\%$ and VoxPopuli $-0.47\%$ relative). The reductions are concentrated on intrinsically hard inputs. On accented and code-switched speech (ASCEND), \method{} achieves a $16.0\%$ relative CER reduction. Across 30 FLEURS languages ($23{,}049$ utterances), the multilingual WER decreases uniformly across resource tiers, confirming that the adapter generalizes without overfitting. Dynamic halting preserves most of the clean-set reduction at a fraction of the compute, skipping roughly half of all utterances at the entry gate. Our results show that a small, carefully chosen activation set can switch on test-time scaling inside a frozen ASR model without corrupting the model itself, converting fixed per-utterance compute into input-dependent compute where it is most needed.
\end{abstract}

\section{Introduction}

End-to-end automatic speech recognition (ASR) models map audio directly to text in a single forward pass~\cite{radford2023robust,peng2025owsm}. This design is simple and effective. However, it forces a single left-to-right decoder to handle acoustic disambiguation, language modeling, and error correction at the same time.

Recent work has shown that allocating extra computation at inference time can substantially improve model performance. This paradigm is broadly known as test-time scaling, and includes explicit chain-of-thought reasoning~\cite{wei2022chain}, parallel sampling, and iterative refinement. Within this line, latent test-time scaling has drawn growing attention. For example, Coconut replaces discrete chain-of-thought tokens with continuous hidden states that are iteratively fed back into the model~\cite{hao2024training}. Quiet-STaR trains models to generate and exploit implicit thoughts to improve next-token prediction~\cite{zelikman2024quiet}. Pause tokens provide extra compute steps without requiring explicit intermediate text~\cite{goyal2024think}. Overall, these findings suggest that latent compute scaling is particularly beneficial on challenging inputs.

Motivated by this, we ask a simple question. Can an ASR decoder benefit from latent test-time scaling before it commits to a transcript?

Applying continuous latent scaling to ASR introduces two challenges absent in standard NLP tasks. First, ASR is a direct transcription task with no intermediate reasoning trajectory to distill, unlike Coconut and Quiet-STaR which rely on explicit chain-of-thought rationales. The model must discover how to use extra latent computation entirely unsupervised. Second, modern ASR backbones are typically kept frozen for efficiency and zero-shot generalization. Standard parameter-efficient methods (LoRA, prompt tuning, prefix tuning) follow this regime~\cite{hu2022lora,lester2021power,li2021prefix}, but recent test-time scaling techniques often fine-tune the entire backbone, which inevitably triggers catastrophic forgetting. Conversely, directly injecting unconstrained continuous states into a frozen backbone causes a geometric mismatch that pushes the decoder out of its native vocabulary distribution and leads to representation collapse.

We propose \method{}, a parameter-efficient method that adds two small trainable modules (a Latent Adapter and a Value Head) to a frozen ASR backbone. The Latent Adapter refines the embeddings of $N$ latent prefix positions through bounded recurrent updates; the Value Head predicts per-utterance utility and halts the loop early. Both modules consume only decoder hidden states; the encoder and decoder weights are never updated.

Our contributions are as follows:
\begin{itemize}
    \item We introduce continuous latent test-time scaling into frozen ASR backbones. We propose a soft-gated additive injection strategy that produces bounded updates and constructs scale-invariant compute states without triggering representation collapse. A component ablation confirms each of the three stabilization mechanisms is individually critical, with removal costs of $+3$ to $+47$~pp WER.
    \item We design a Value Head that predicts per-utterance utility and halts computation when extra steps will not help. Dynamic halting cuts the average per-utterance step count from $4$ to $0.71$--$1.24$ while preserving the WER gain. On harder data (ASCEND accented speech) the same policy allocates substantially more compute ($34.1\%$ of utterances trigger the full $N{=}4$ budget), demonstrating input-dependent compute allocation.
    \item We demonstrate \emph{minimal-data activation}: a deliberately small, diverse 500-utterance training set is sufficient to \emph{activate} latent test-time scaling on a frozen backbone, but too small to \emph{adapt} (and thereby corrupt) the backbone itself. Standard adaptation methods of comparable scale instead all regress. At this optimum, \method{} yields a $-2.54\%$ relative WER on FLEURS, a $-0.47\%$ relative WER on VoxPopuli, a $-16.0\%$ relative CER on ASCEND accented/code-switched speech, and directional reductions across a 23{,}049-utterance multilingual evaluation spanning 30 languages, all with only ${\sim}4.0$M trainable parameters (${\sim}0.67\%$ overhead).
\end{itemize}

\section{Related Work}
\label{sec:related}

\textbf{Latent test-time scaling.}\quad Beyond explicit chain-of-thought reasoning~\cite{wei2022chain}, recent work injects \emph{continuous} computation into language models. Coconut~\cite{hao2024training} feeds last-layer hidden states back into the model in place of discrete tokens, allowing reasoning to proceed in a continuous embedding space. Quiet-STaR~\cite{zelikman2024quiet} trains models to generate and exploit implicit thoughts that improve next-token prediction without explicit rationale supervision. Pause tokens~\cite{goyal2024think} insert filler positions that grant the model extra forward compute without producing intermediate text. These methods primarily target language modeling and reasoning benchmarks where intermediate rationales either exist or can be distilled from teacher models. \method{} extends the latent-scaling principle to direct speech-to-text transcription, where the target sequence is the only supervision signal and the backbone must remain frozen to preserve broad acoustic generalization.

\textbf{Adaptive computation.}\quad Spending variable compute per input has a long history. Adaptive Computation Time (ACT)~\cite{graves2016adaptive} learns a halting probability per step inside an RNN, terminating when accumulated halting mass exceeds a threshold. PonderNet~\cite{banino2021pondernet} reformulates this with a probabilistic framework and explicit regularization on the expected pondering steps. Early-exit transformers~\cite{schuster2022confident,xin2020deebert} cut inference cost by exiting at intermediate layers when token-level confidence is high. \method{}'s Value Head sits in this lineage but differs along two axes. First, it operates over a small fixed budget of \emph{latent prefix} positions rather than over network depth or over output sequence length. Second, its halting signal is the predicted \emph{latent-vs-baseline accuracy gain}, not a confidence proxy or a learned ponder cost. The signal is therefore directly aligned with the deployment objective, rather than a heuristic surrogate.

\textbf{Parameter-efficient adaptation.}\quad LoRA~\cite{hu2022lora}, prompt tuning~\cite{lester2021power}, and prefix tuning~\cite{li2021prefix} keep a backbone frozen while inserting low-rank weight updates or learnable surface prompts. As shown in Table~\ref{tab:main}, under our 500-utterance low-resource regime these gradient-based input or weight perturbations consistently \emph{increase} WER on a strong zero-shot ASR backbone, by double-digit relative percentages for full fine-tuning and LoRA, and catastrophically for prompt tuning. \method{} differs in two ways. First, it leaves both backbone weights and surface prompts untouched. Second, it does not minimize the input-output discrepancy; it instead introduces additional \emph{computation} along a bounded latent trajectory governed by a learned halting policy.

\textbf{Robust pre-trained ASR.}\quad Foundation ASR models such as Whisper~\cite{radford2023robust}, OWSM~\cite{peng2025owsm}, and Qwen3-ASR~\cite{shi2026qwen3asr} achieve strong zero-shot transcription via massive multilingual pre-training. Their generalization, however, is brittle: even small downstream gradient updates can erase it (Section~\ref{sec:exp_setup}). To our knowledge, \method{} is the first method to add a learnable test-time refinement loop on top of such a backbone without modifying any of its parameters, recovering WER reductions in regimes where conventional adaptation actively regresses.

\section{Method}

\subsection{Problem Setup}
\label{sec:setup}

We consider an end-to-end ASR pipeline with a frozen acoustic encoder and a frozen autoregressive text decoder. Given an utterance $a$, the encoder produces acoustic states $\mathbf{Z}=\mathrm{Enc}(a)$, and the decoder generates transcript tokens conditioned on $\mathbf{Z}$ and a text prefix. \method{} keeps both the encoder and the decoder fully frozen and intervenes \emph{only} on the decoder's input prefix: it inserts $N$ \emph{latent positions} between the system prompt and the transcript target. These positions are non-text continuous slots used solely as additional computation context; no transcript token is emitted at them. We use $N$ both as the number of latent positions and as the maximum number of refinement steps: refinement step $k$ updates the embedding at latent position $k$.

\subsection{Overview}

\begin{figure*}[t]
\centering
\includegraphics[width=\linewidth]{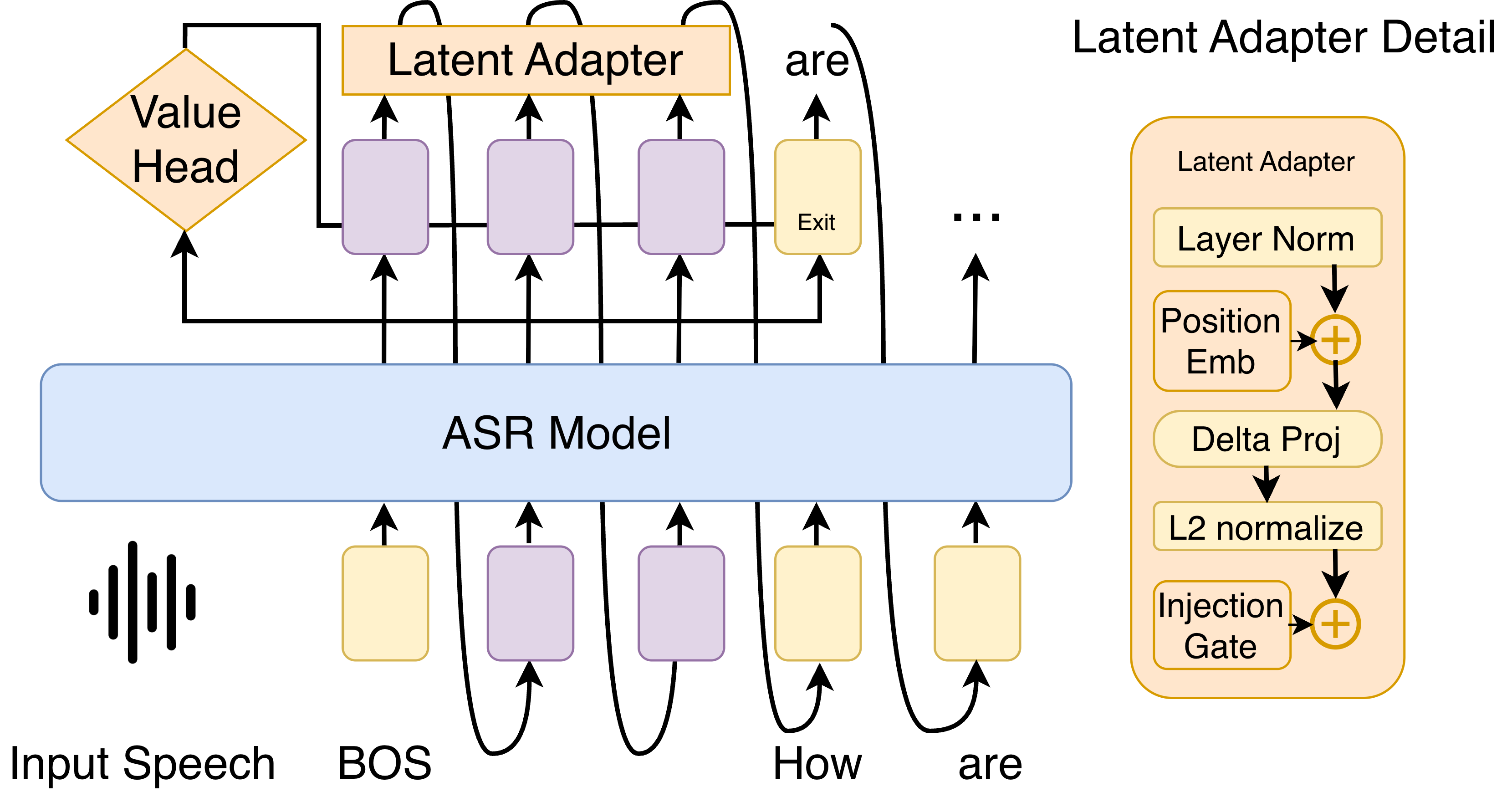}
\caption{Architecture of \method{}. The ASR backbone (blue) remains \textbf{frozen}. We add a trainable \textbf{Latent Adapter} (orange) that forms a recurrent compute loop. At step $k$, it maps the decoder hidden state at latent position $k$ to a bounded update $\vdelta_k$ that is added to the latent embedding via a sigmoid gate. A \textbf{Value Head} (diamond) reads each latent state to predict a utility score $v_k$, enabling the model to halt dynamically when further computation is unnecessary.}
\label{fig:arch}
\end{figure*}

\method{} adds two small trainable modules to the frozen backbone (Fig.~\ref{fig:arch}): a \textbf{Latent Adapter} that refines the embeddings at the $N$ latent positions through a bounded recurrent loop, and a \textbf{Value Head} that decides how many of those refinement steps to actually run. Both modules consume only decoder hidden states; neither modifies the backbone or the encoder. One difficulty drives the whole design. Feeding an arbitrary continuous vector into a frozen decoder pushes its input off the distribution seen during pre-training and quickly destabilizes it. The Latent Adapter therefore does not inject free vectors; it produces small, bounded, gated updates around a fixed anchor embedding, controlled by three stabilization mechanisms that we detail in Section~\ref{sec:stability}. Sections~\ref{sec:refinement}--\ref{sec:valuehead} describe each component in turn.

\subsection{Latent Prefix Refinement}
\label{sec:refinement}

The Latent Adapter computes a bounded update at each refinement step. Let $\mathbf{h}_0$ denote the decoder's last-layer hidden state at the final latent position before any refinement (the \emph{acoustic anchor}); we $L_2$-normalize it before downstream use so the loop starts from a unit-magnitude state. At refinement step $k\in\{1,\dots,N\}$, the frozen decoder reads $\mathbf{Z}$ and the current latent prefix; we extract the last-layer hidden state $\mathbf{h}_k$ at latent position $k$. The adapter, time-conditioned by a learned step embedding $\mathbf{e}_k$, projects $\mathbf{h}_k$ into the embedding space, $L_2$-normalizes the projection, and rescales by a learned scalar $s_k$ to produce a step-bounded delta:
\begin{equation}
    \vdelta_k = s_k \cdot \frac{\text{DeltaProj}(\mathbf{h}_k)}{\lVert\text{DeltaProj}(\mathbf{h}_k)\rVert + \epsilon}, \quad \epsilon{=}10^{-6}.
    \label{eq:delta}
\end{equation}
The $L_2$-normalization makes the update direction scale-invariant; the per-step scale $s_k$ controls how far each step moves.

\subsection{Stable Injection into a Frozen Decoder}
\label{sec:injection}
\label{sec:stability}

A frozen decoder only interprets inputs that resemble the token
embeddings it saw during pre-training. Injecting an arbitrary
continuous vector pushes the input off this embedding manifold and
destabilizes the decoder. LatentASR keeps the injection on-manifold by
constraining three distinct properties of the update: how far each step
moves the state, whether the step is applied, and where the resulting
embedding sits. Each property has its own failure mode, so the three
mechanisms are complementary rather than redundant.

\begin{itemize}
    \item \textbf{Bounded delta.} The per-step scale $s_k$ and the
    $L_2$-normalization in Eq.~\ref{eq:delta} bound the magnitude of
    every update, so no single step can grow without limit. Without this
    bound the update magnitude grows without constraint and pushes the
    latent state off the manifold within a few steps.

    \item \textbf{Sigmoid gate.} A learned gate decides how much of each
    update to apply, and can veto a refinement on positions that already
    look correct:
    \begin{equation}
        g_k = \sigma\big(\mathbf{W}_g [\mathbf{e}_{\texttt{LT}}; \vdelta_k]\big),
        \label{eq:gate}
    \end{equation}
    with $\mathbf{W}_g$ zero-initialized so $g_k{=}0.5$ at the start of
    training. The model begins neutral and learns to drive the gate
    toward $0$ or $1$. Without the gate every step applies the full
    update and over-corrects positions that are already right.

    \item \textbf{Fixed-embedding anchor.} Every latent position is
    anchored to the same fixed token embedding $\mathbf{e}_{\texttt{LT}}$
    and modified only by an additive residual,
    \begin{equation}
        \mathbf{x}_k = \mathbf{e}_{\texttt{LT}} + g_k \cdot \vdelta_k,
        \label{eq:inject}
    \end{equation}
    so the injected embedding stays in a small ball around a real
    vocabulary embedding the decoder already interprets. A bounded step
    size alone does not guarantee this. It limits per-step displacement
    but not absolute location, so without the anchor the latent state can
    still drift away from any real embedding.
\end{itemize}

Together, Eqs.~\ref{eq:delta}--\ref{eq:inject} form a gated residual
around a fixed anchor. Closing the gate recovers
$\mathbf{e}_{\texttt{LT}}$ exactly, so the adapter only learns a small,
optionally-zero correction rather than a full continuous input from
scratch. The component ablation in Section~\ref{sec:ablation} confirms
the three mechanisms are not interchangeable: removing any one degrades
WER substantially, by $+3$ to $+47$ pp.

\subsection{Utility Prediction and Dynamic Halting}
\label{sec:valuehead}

Latent scaling reliably helps some utterances and merely costs compute on others. We therefore need a per-utterance gate that predicts, before any compute is spent, whether further latent steps will help. The Value Head plays this role.

\textbf{Design rationale.}\quad The signal we want is ``would extra latent compute help this utterance?'', which is most directly answered by comparing the latent path against the frozen baseline on the same training utterance, a self-distilled target whose value is the latent-vs-baseline accuracy gain $\Delta$ on supervised transcript positions. Two practical issues shape the rest of the design. \emph{(1)} As the latent path improves during training, $\Delta$ becomes positive on most utterances and the regression target loses its negative-utility regime. We expose a \emph{forced-negative sampling} knob $p_{\text{neg}}$ that, with probability $p_{\text{neg}}$ per minibatch, flips $y \leftarrow -|y|$, providing a tunable counterweight against this drift. We default to $p_{\text{neg}}{=}0.3$ for conservative calibration, and treat it as an optional calibration knob rather than a strictly required safeguard. \emph{(2)} On extreme OOD utterances where both passes attain zero token accuracy, $\Delta$ is undefined precisely where the gate most needs to learn that compute will not help; we restore signal there with a clamped ($[-2, 2]$) cross-entropy fallback at a smaller temperature $\gamma{=}0.5$.

\textbf{Implementation.}\quad The Value Head is a single linear layer over each latent hidden state, $v_k = \tanh(\text{ValueHead}(\mathbf{h}_k)) \in [-1, 1]$, so larger $v_k$ predicts a more negative $\Delta$WER. On non-degenerate utterances the target is the tanh-squashed accuracy gain
\begin{equation}
    y = \alpha \cdot \tanh(\gamma \cdot \Delta),
    \label{eq:value_target}
\end{equation}
with $\alpha{=}0.9$ (label-smoothing boundary) and $\gamma{=}3.0$ (gain temperature).

\textbf{Two-stage halting at inference.}\quad $v_0$ on the initial anchor gates the entire loop: if $v_0 < \theta$, all $N$ steps are skipped and the utterance falls back to the frozen Baseline. Otherwise, after each step $k\geq 1$, we re-evaluate $v_k$ and break if it drops below $\theta$. Latent compute is thus allocated only to utterances the Value Head predicts will benefit.

\subsection{Training}
\label{sec:training}

Three losses train the adapter and Value Head while the backbone stays frozen:
\begin{equation}
    \mathcal{L} = \mathcal{L}_{\text{CE}} + w_{\text{cyc}}(t) \cdot \mathcal{L}_{\text{cyc}} + w_{\text{val}} \cdot \mathcal{L}_{\text{val}}.
    \label{eq:loss_total}
\end{equation}
$\mathcal{L}_{\text{CE}}$ is cross-entropy on the transcript tokens (latent positions are masked out). $\mathcal{L}_{\text{val}}$ is the mean-squared error between the Value Head output and the target in Eq.~\ref{eq:value_target}. $\mathcal{L}_{\text{cyc}}$ is a trajectory regularizer
\begin{equation}
\mathcal{L}_{\text{cyc}} = \tfrac{\beta}{N}\!\sum_{k=1}^{N}\!\lVert\mathbf{h}_k {-} \mathbf{h}_0\rVert_2^2 + \tfrac{1{-}\beta}{N{-}1}\!\sum_{k=1}^{N-1}\!\lVert\mathbf{h}_{k+1} {-} \mathbf{h}_k\rVert_2^2,
\label{eq:loss_cyc}
\end{equation}
that anchors the latent trajectory near $\mathbf{h}_0$ and damps abrupt step-to-step jumps, with $\beta{=}0.3$ favoring smooth refinement over rigid attachment. The cycle weight $w_{\text{cyc}}(t)$ decays linearly to zero, so the trajectory is constrained early when the geometry is unstable and is gradually freed to specialize. To prevent over-reliance on the latent path, we additionally drop the entire latent loop with probability $p_{\text{drop}}$ during training and add Gaussian noise of std $\sigma$ to the initial anchor.

\section{Experiments}
\subsection{Experiment Setup}
\label{sec:exp_setup}
\textbf{Training Setup.}\quad We use Qwen3-ASR-0.6B~\cite{shi2026qwen3asr} as the base model and keep all its parameters frozen. We train for 10 epochs with effective batch size $16$ on a single NVIDIA RTX 5090. We set $N{=}4$ and use AdamW with decoupled learning rates: $10^{-4}$ for the Latent Adapter and Value Head, and $5 \times 10^{-5}$ for step scales (initialized at $0.2$, capped at $3.0$). We use $w_{\text{cyc}}{=}0.1$ (linearly decayed), $w_{\text{val}}{=}3.0$, $p_{\text{drop}}{=}0.15$, input noise $\sigma{=}0.5$, and $p_{\text{neg}}{=}0.3$. With this configuration, the trainable components add only $\sim 4.0\text{M}$ parameters, yielding minimal overhead over the frozen backbone.

\textbf{Dataset: Minimal-Data Activation.}\quad We frame our training setup as a deliberate \emph{minimal-data activation} strategy. Large-scale fine-tuning erases the frozen backbone's zero-shot generalization, yet the latent adapter still needs supervision to learn \emph{when} extra test-time computation helps. A small but acoustically diverse training set supplies this signal without exerting enough gradient pressure to corrupt the backbone. Section~\ref{sec:ablation} validates this with an activation-size sweep (100 to 800 utterances) that places the optimum at 500.

We construct a 500-utterance training mixture from seven ASR corpora. The mixture is dominated by Common Voice 16.0~\cite{commonvoice:2020}, supplemented by FLEURS~\cite{fleurs2022arxiv} and VoxPopuli~\cite{wang-etal-2021-voxpopuli}, and regularized with challenging tail samples from LibriSpeech~\cite{panayotov2015librispeech}, GigaSpeech~\cite{GigaSpeech2021}, The People's Speech~\cite{DBLP:journals/corr/abs-2111-09344}, and ASCEND~\cite{lovenia2022ascend}. It is strongly multilingual, covering 100+ language codes and intentionally including fragmented English dialect tags (e.g., \texttt{en}, \texttt{en\_us}, \texttt{en\_accented}). Two design principles drive the construction. First, the high diversity prevents the adapter from overfitting to any single dataset-locale combination, forcing it to learn generalizable latent representations. Second, the small scale keeps gradient pressure on the frozen backbone minimal. We refer to this principle as \emph{minimal-data activation}: the small, varied training set is enough to \emph{activate} latent test-time scaling, but too small to \emph{adapt} the frozen backbone off its pre-trained manifold. The 500-utterance activation set is sampled from an 811-utterance mixture with seed 42. We evaluate on the standardized FLEURS (\texttt{en\_us}) test set, VoxPopuli (\texttt{en}), and ASCEND.

\textbf{Baselines.}\quad We benchmark \method{} against the frozen Qwen3-ASR-0.6B baseline, full fine-tuning, LoRA~\cite{hu2022lora} (rank 16), and prompt tuning~\cite{lester2021power} (4 soft tokens). To match the activation regime, training is restricted to the same 500-utterance diverse mixture for all methods. This tests adaptation to complex acoustics without triggering representation collapse. All models are trained for 10 epochs using AdamW (weight decay 0.01, max norm 1.0). Learning rates follow original recipes: $1 \times 10^{-5}$ for full fine-tuning, $1 \times 10^{-4}$ for LoRA, and $5 \times 10^{-4}$ for prompt tuning. The latter two utilize a 10\% linear warmup with cosine decay. To ensure fair comparison, \method{}, LoRA, and prompt tuning operate with the ${\sim}600$M backbone entirely frozen. Full fine-tuning was rerun with micro-batch 4 and gradient accumulation 4 after the larger micro-batch OOMed, preserving the same effective batch size.

\textbf{Checkpoint selection.}\quad For clarity, Table~\ref{tab:main} uses the canonical checkpoint selected from the activation-size sweep. The ablation tables, and the Pareto curves in Fig.~\ref{fig:pareto}, use independently retrained same-protocol checkpoints, because each ablation changes the training configuration. Their purpose is to test mechanism sensitivity, not to provide an exact rerun of the canonical Table~\ref{tab:main} checkpoint. Consequently, small absolute differences between the full-model rows in the ablation tables (and Fig.~\ref{fig:pareto}) and Table~\ref{tab:main} should be interpreted as run-to-run variation under the same protocol.

\subsection{Main Results}
\label{sec:main_results}

Table~\ref{tab:main} reports performance on FLEURS (\texttt{en\_us}) and VoxPopuli (\texttt{en}). Throughout, $\Delta$WER is the relative change over the frozen Baseline (negative = improvement). Across both test sets, every conventional adaptation method we test increases WER under the 500-utterance training regime. LoRA regresses by $+32.0\%$ relative on FLEURS, full fine-tuning by $+13.4\%$, and prompt tuning collapses outright (FLEURS WER rises to $84.80\%$). Prompt tuning's collapse shows that even small input-distribution updates can push a strong frozen ASR backbone out of its operating regime entirely.

In contrast, \method{} is the only tested method that reduces WER on both benchmarks. On FLEURS, it reduces WER from $4.900\%$ to $4.776\%$ ($-2.54\%$ relative) and CER from $2.326\%$ to $2.238\%$ ($-3.78\%$ relative). On VoxPopuli, it reduces WER from $9.038\%$ to $8.995\%$ ($-0.47\%$ relative). These results support our \emph{minimal-data activation} hypothesis: the same 500 noisy and diverse samples that destabilize standard adaptation are sufficient to activate latent test-time scaling on a frozen backbone. Crucially, the adapter provides per-utterance refinement space without overfitting. The backbone remains untouched and its zero-shot generalization is fully preserved.

\begin{table}[ht]
  \caption{WER and CER on the FLEURS (\texttt{en\_us}, 647 utts) and VoxPopuli (\texttt{en}, 1{,}842 utts) test splits. \textbf{Rel. $\boldsymbol{\Delta}$WER (\%)} is the relative WER change vs.\ the frozen Baseline. \textbf{Negative = WER reduction (improvement); positive = WER increase (regression).} Prompt tuning collapses outright under the 500-utterance regime.}
  \label{tab:main}
  \centering
  \resizebox{\columnwidth}{!}{
  \begin{tabular}{l | c c c | c c c}
    \toprule
    \multirow{2}{*}{\textbf{Model}} & \multicolumn{3}{c|}{\textbf{FLEURS}} & \multicolumn{3}{c}{\textbf{VoxPopuli}} \\
    \cmidrule(lr){2-4} \cmidrule(lr){5-7}
     & \textbf{WER (\%)} & \textbf{CER (\%)} & \textbf{Rel.\ $\Delta$WER (\%)} & \textbf{WER (\%)} & \textbf{CER (\%)} & \textbf{Rel.\ $\Delta$WER (\%)} \\
    \midrule
    Baseline               & 4.900 & 2.326 & -- & 9.038 & 5.900 & -- \\
    \quad + Full Fine-Tuning & 5.556 & 2.481 & $+13.38$ & 9.485 & 6.176 & $+4.96$ \\
    \quad + Prompt Tuning  & 84.803 & 82.468 & $+1630.70$ & 88.507 & 87.901 & $+879.33$ \\
    \quad + LoRA (rank 16) & 6.467 & 2.944 & $+31.97$ & 10.212 & 6.536 & $+13.00$ \\
    \midrule
    \quad + \method{} ($N{=}4$, $\theta{=}0$) & \textbf{4.776} & \textbf{2.238} & $\boldsymbol{-2.54}$ & \textbf{8.995} & \textbf{5.869} & $\boldsymbol{-0.47}$ \\
    \bottomrule
  \end{tabular}
  }
\end{table}

\subsection{Multilingual and Cross-Domain Generalization}
\label{sec:multilingual}

To verify that \method{} preserves the broad multilingual capability of the frozen backbone, we evaluate on a 23{,}049-utterance multilingual benchmark spanning the FLEURS test sets of 30 languages.

\textbf{Multilingual aggregate.}\quad Table~\ref{tab:multilingual_agg} reports aggregate WER on FLEURS 30 (23{,}049 utterances) and its three resource-tier partitions: \textbf{core12} (12 high-resource languages including \texttt{en\_us}), \textbf{+8} (8 medium-resource), and \textbf{+10} (10 low-resource). \method{} reduces WER across all three tiers, with the largest relative improvement on the high-resource tier ($-0.26\%$ rel). The uniformly negative $\Delta$WER across tiers confirms that a 500-utterance activation set generalizes without overfitting. The adapter does not specialize to any particular resource tier or language family.

\begin{table}[ht]
  \caption{Multilingual aggregate WER on FLEURS 30 and its resource-tier partitions. \textbf{Rel.\ $\boldsymbol{\Delta}$WER (\%)} is computed against the frozen Baseline; negative = WER reduction = improvement.}
  \label{tab:multilingual_agg}
  \centering
  \resizebox{\columnwidth}{!}{
  \begin{tabular}{l r c c c}
    \toprule
    \textbf{Subset} & \textbf{\#utts} & \textbf{Baseline WER (\%)} & \textbf{\method{} WER (\%)} & \textbf{Rel.\ $\Delta$WER (\%)} \\
    \midrule
    FLEURS 30 (all)         & 23{,}049 & 32.65 & \textbf{32.57} & $-0.23$ \\
    \quad core12            &  8{,}876 & 31.71 & \textbf{31.63} & $-0.26$ \\
    \quad +8                &  5{,}597 & 17.48 & \textbf{17.45} & $-0.19$ \\
    \quad +10               &  8{,}576 & 43.52 & \textbf{43.42} & $-0.23$ \\
    \bottomrule
  \end{tabular}
  }
\end{table}

\textbf{Per-language reductions.}\quad Table~\ref{tab:per_language} reports two character-based FLEURS languages, for which CER is the appropriate metric. The two cases sharpen the difficulty-conditional reading. Thai and Japanese start from almost the same baseline CER ($9.53\%$ and $10.35\%$). The latent loop still treats them differently. It reduces Thai CER by $-1.31\%$ relative, but leaves Japanese exactly neutral ($0.00\%$). The effect is therefore not a function of the baseline error rate. It instead tracks whether the residual errors are systematically correctable. On Thai these are script and segmentation normalization errors; on Japanese the residual errors are not of a correctable kind. Neither language regresses. This matches the frozen-backbone design. The adapter refines where it helps and stays inert otherwise, so broad multilingual capability is preserved while targeted reductions accumulate where the decoder's residual errors are systematic.

\begin{table}[ht]
  \caption{Per-language results on character-based FLEURS languages. Negative Rel.\ $\boldsymbol{\Delta}$CER = error reduction = improvement.}
  \label{tab:per_language}
  \centering
  \resizebox{\columnwidth}{!}{
  \begin{tabular}{l r c c c}
    \toprule
    \textbf{Language (FLEURS code)} & \textbf{\#utts} & \textbf{Baseline CER (\%)} & \textbf{\method{} CER (\%)} & \textbf{Rel.\ $\Delta$CER (\%)} \\
    \midrule
    Japanese (ja\_jp)        &   650 & 10.35 & 10.35 & $\phantom{-}0.00$ \\
    Thai (th\_th)            & 1{,}021 &  9.53 & \textbf{ 9.41} & $\boldsymbol{-1.31}$ \\
    \bottomrule
  \end{tabular}
  }
\end{table}

\subsection{Robustness Study}
\label{sec:robustness}
To test whether \method{} preserves the frozen backbone's generalization under acoustic shifts, we evaluate on additive Gaussian noise at SNR $= 0$~dB and on heavily accented / code-switched speech (Table~\ref{tab:robustness}).

\textbf{Zero-SNR Gaussian noise.}\quad On the two main benchmarks under SNR$=0$~dB, \method{} reduces WER on both: FLEURS by $-1.32\%$ relative ($29.81 \to 29.42$) and VoxPopuli by $-1.02\%$ relative ($19.71 \to 19.51$). The latent loop is therefore robust on read- and lecture-style speech under heavy noise.

\textbf{Accented and code-switched speech.}\quad We further evaluate on ASCEND, which contains strong regional accents and frequent code-switching. Word-level metrics can be unreliable on such data; we therefore use CER, following~\cite{lovenia2022ascend}. \method{} achieves a $-16.02\%$ relative CER reduction ($57.81 \to 48.55$). This large reduction demonstrates that the latent loop provides substantial refinement capacity on acoustically challenging speech. Section~\ref{sec:ablation} shows that on ASCEND the Value Head allocates substantially more compute than on read-speech benchmarks ($34.1\%$ of utterances trigger the full $N{=}4$ budget), confirming input-dependent compute allocation on intrinsically harder utterances.

\begin{table}[ht]
  \caption{Robustness evaluations on noisy and accented conditions. \textbf{Rel.\ $\boldsymbol{\Delta}$ (\%)} is the relative WER (or CER) change vs.\ the Baseline; negative = error reduction = improvement.}
  \label{tab:robustness}
  \centering
  \resizebox{\columnwidth}{!}{
  \begin{tabular}{l l c c c c}
    \toprule
    \textbf{Dataset} & \textbf{Condition} & \textbf{Metric} & \textbf{Baseline (\%)} & \textbf{\method{} (\%)} & \textbf{Rel.\ $\Delta$ (\%)} \\
    \midrule
    FLEURS    & SNR$=0$~dB & WER & 29.81 & \textbf{29.42} & $-1.32$ \\
    VoxPopuli & SNR$=0$~dB & WER & 19.71 & \textbf{19.51} & $-1.02$ \\
    \midrule
    ASCEND    & accented   & CER & 57.81 & \textbf{48.55} & $\boldsymbol{-16.02}$ \\
    \bottomrule
  \end{tabular}
  }
\end{table}

\textbf{Stress Broad Suite.}\quad
To test whether the latent loop activates when the acoustic conditions become substantially harder, we evaluate the same checkpoint on six clean-speech corpora corrupted with additive Gaussian noise at SNR$=0$~dB: AMI~\cite{carletta2005ami} (IHM+SDM), Common Voice (en)~\cite{commonvoice:2020}, GigaSpeech~\cite{GigaSpeech2021}, LibriSpeech (clean+other)~\cite{panayotov2015librispeech}, The People's Speech~\cite{DBLP:journals/corr/abs-2111-09344}, and TEDLIUM~\cite{rousseau2012tedlium}. Each configuration is capped at 500 utterances where available ($3{,}887$ utterances in total). Table~\ref{tab:stress_broad} reports per-corpus WER. The utterance-weighted aggregate WER drops from $36.84\%$ to $36.27\%$ ($-0.58$~pp, $-1.57\%$ rel). The largest reductions appear on the most acoustically degraded corpora (AMI $-2.91\%$ rel, TEDLIUM $-3.76\%$ rel), consistent with the difficulty-conditional design: the latent loop contributes most where the frozen decoder leaves the largest correctable residual.

\begin{table}[ht]
  \caption{Stress broad suite at SNR$=0$~dB. Weighted aggregate WER drops by $-0.58$~pp ($-1.57\%$ rel).}
  \label{tab:stress_broad}
  \centering
  \resizebox{\columnwidth}{!}{
  \begin{tabular}{l r c c c c}
    \toprule
    \textbf{Dataset} & \textbf{\#utts} & \textbf{Baseline WER (\%)} & \textbf{\method{} WER (\%)} & \textbf{$\Delta$WER (pp)} & \textbf{Rel.\ $\Delta$WER (\%)} \\
    \midrule
    AMI (IHM+SDM) $0$~dB              &   996 & 77.784 & \textbf{75.518} & $\boldsymbol{-2.266}$ & $\boldsymbol{-2.91}$ \\
    Common Voice (en) $0$~dB          &   500 & 31.617 & 31.936 & $+0.319$ & $+1.01$ \\
    GigaSpeech (test) $0$~dB          &   391 & 20.939 & \textbf{20.875} & $-0.064$ & $-0.30$ \\
    LibriSpeech (clean+other) $0$~dB  & 1{,}000 & 18.471 & 18.584 & $+0.113$ & $+0.61$ \\
    People's Speech $0$~dB            &   500 & 31.687 & 31.771 & $+0.084$ & $+0.26$ \\
    TEDLIUM (release 1) $0$~dB        &   500 & 14.849 & \textbf{14.291} & $-0.558$ & $-3.76$ \\
    \midrule
    \textbf{Weighted aggregate}       & \textbf{3{,}887} & \textbf{36.843} & \textbf{36.265} & $\boldsymbol{-0.578}$ & $\boldsymbol{-1.57}$ \\
    \bottomrule
  \end{tabular}
  }
\end{table}

\subsection{Ablation Study}
\label{sec:ablation}

We ablate the core design choices of \method{}: (i) the halting threshold $\theta$ and the latent budget $N$; (ii) the three stabilization mechanisms in Section~\ref{sec:stability}; and (iii) the size of the activation set.

\begin{figure}[ht]
\centering
\begin{tikzpicture}
\begin{axis}[
    width=0.95\columnwidth,
    height=5.0cm,
    xlabel={Average latent steps per utterance},
    ylabel={$\Delta$WER (pp, $\downarrow$ better)},
    xmin=-0.25, xmax=4.4,
    ymin=-0.10, ymax=0.04,
    ymajorgrids=true,
    xmajorgrids=true,
    grid style={dashed, gray!30},
    legend pos=north east,
    legend style={font=\scriptsize, draw=gray!50, fill=white, fill opacity=0.9, text opacity=1},
    legend cell align=left,
    label style={font=\small},
    tick label style={font=\scriptsize},
    every axis plot post/.append style={thick},
]
\addplot[gray!60, dashed, thin, mark=none, forget plot]
    coordinates {(-0.25, 0) (4.4, 0)};

\addplot[blue!75!black, mark=*, mark size=2.0pt, mark options={fill=blue!75!black}]
    coordinates {
    (0.00,  0.000)
    (1.24, -0.041)
    (1.37, -0.041)
    (4.00, -0.062)
};
\addlegendentry{FLEURS (\texttt{en\_us})}

\addplot[red!75!black, mark=square*, mark size=2.0pt, mark options={fill=red!75!black}]
    coordinates {
    (0.00,  0.000)
    (0.71, -0.063)
    (0.84, -0.070)
    (4.00, -0.014)
};
\addlegendentry{VoxPopuli (\texttt{en})}

\addplot[blue!75!black, mark=o, mark size=5pt, line width=1.2pt, only marks, forget plot]
    coordinates {(1.24, -0.041)};
\addplot[red!75!black, mark=o, mark size=5pt, line width=1.2pt, only marks, forget plot]
    coordinates {(0.71, -0.063)};

\node[font=\scriptsize, anchor=west, fill=white, fill opacity=0.85, text opacity=1, inner sep=1pt]
    at (axis cs:1.8, -0.085) {deployed: $\theta{=}0$};
\draw[->, gray!70, thin]
    (axis cs:1.8, -0.085) -- (axis cs:1.27, -0.043);
\draw[->, gray!70, thin]
    (axis cs:1.8, -0.085) -- (axis cs:0.78, -0.063);

\end{axis}
\end{tikzpicture}
\caption{$\Delta$WER--compute Pareto curves from sweeping $\theta$ ($\downarrow$ better). FLEURS rewards more latent steps; VoxPopuli reaches its minimum at low compute. The deployed $\theta{=}0$ (hollow circles) sits near the knee of both frontiers. The point at $(0,0)$ is the saturated regime ($\theta{\geq}0.2$), where all utterances skip. These curves use the same-protocol sweep checkpoint, so the absolute $\Delta$WER differs from the canonical Table~\ref{tab:main} checkpoint (Section~\ref{sec:exp_setup}).}
\label{fig:pareto}
\end{figure}
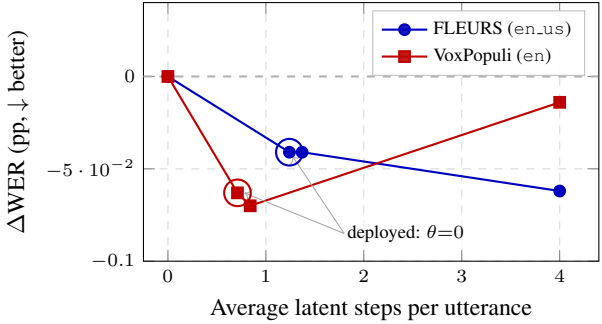

\textbf{$\boldsymbol{\Delta}$WER--Compute Pareto.}\quad
Fig.~\ref{fig:pareto} sweeps the halting threshold $\theta$, tracing $\Delta$WER against the average per-utterance step count (lower is better). The two datasets sit at different operating points: FLEURS rewards more latent steps (down to $\Delta$WER $=-0.062$~pp at $4.00$ steps), whereas VoxPopuli reaches its minimum below one step. The deployed setting ($\theta{=}0.0$) sits near the knee on both curves, capturing roughly $66\%$ (FLEURS) and $90\%$ (VoxPopuli) of the available reduction at a fraction of full compute. A static budget cannot recover this point: any uniform $N$ either under-allocates on FLEURS or over-spends on VoxPopuli. At $\theta{\geq}0.2$ all utterances skip and \method{} collapses exactly to the Baseline, confirming that the reductions arise from learned latent refinement rather than from a decoding-pipeline difference.

\textbf{Compute Allocation.}\quad
Table~\ref{tab:halt_dist} reports the per-step distribution at $\theta{=}0.0$ across three difficulty levels. The skip rate falls cleanly with difficulty ($60.2\%$ VoxPopuli, $47.0\%$ FLEURS, $22.3\%$ ASCEND), while the full-$N{=}4$ share rises to $34.1\%$ on ASCEND, up to $5.5\times$ the share on the read-speech benchmarks ($6.2\%$ on VoxPopuli, $18.2\%$ on FLEURS). The Value Head therefore acts as a difficulty-aware compute scheduler, allocating more latent steps where the Baseline is weaker.

\begin{table}[th]
  \caption{$N$-step distribution at the deployed setting $\theta{=}0.0$ across three test conditions of varying difficulty.}
  \label{tab:halt_dist}
  \centering
  \resizebox{\columnwidth}{!}{
  \begin{tabular}{l c r r r r}
    \toprule
    \textbf{Dataset} & \textbf{Skip ($\mathbf{N=0}$)} & \textbf{N=1} & \textbf{N=2} & \textbf{N=3} & \textbf{N=4} \\
    \midrule
    VoxPopuli (clean)   & 60.2\% & 24.0\% & 6.7\% & 2.8\% &  6.2\% \\
    FLEURS (clean)      & 47.0\% & 23.2\% & 7.0\% & 4.6\% & 18.2\% \\
    ASCEND (accented)   & 22.3\% & 30.3\% & 8.2\% & 5.1\% & 34.1\% \\
    \bottomrule
  \end{tabular}
  }
\end{table}

\textbf{Component Ablation: Stable Injection Mechanisms.}\quad
The three stabilization mechanisms in Section~\ref{sec:stability} (bounded delta, sigmoid gate, and fixed-embedding anchor) were claimed to target distinct failure modes. Table~\ref{tab:component_ablation} validates that claim by selectively removing each one and retraining under the same regime. The three failure modes are quantitatively distinct and severe. Removing the \emph{bounded delta} (skipping $L_2$-normalization and the per-step scale $s_k$) is catastrophic ($+46.85$~pp WER): the update magnitude grows without constraint and pushes the latent state out of the decoder's input manifold within a few steps. Removing the \emph{fixed-embedding anchor} (replacing $\mathbf{e}_{\texttt{LT}} + g_k\cdot\vdelta_k$ with the raw injection $g_k\cdot\vdelta_k$) costs $+11.33$~pp because the latent token leaves the neighborhood of any real vocabulary embedding. Removing the \emph{sigmoid gate} (fixing $g_k=1$) is the mildest of the three but still costs $+2.99$~pp by forcing every step to apply the full delta even on already-correct positions. The three mechanisms are therefore complementary rather than redundant.

\begin{table}[th]
  \caption{Component ablation on the three stabilization mechanisms (FLEURS \texttt{en\_us}, $\theta{=}0.0$, Baseline WER $4.90\%$). Each row removes one mechanism while keeping the others; all three are individually critical. This suite uses independently retrained same-protocol checkpoints.}
  \label{tab:component_ablation}
  \centering
  \resizebox{\columnwidth}{!}{
  \begin{tabular}{l c c}
    \toprule
    \textbf{Variant} & \textbf{WER (\%)} & \textbf{$\Delta$WER (pp)} \\
    \midrule
    Full \method{} ($N{=}4$, $\theta{=}0.0$) & \textbf{4.86} & $\boldsymbol{-0.04}$ \\
    \quad $-$ bounded delta ($L_2$ + scale $s_k$) & 51.75 & $+46.85$ \\
    \quad $-$ fixed-embedding anchor ($\mathbf{e}_{\texttt{LT}}$ removed) & 16.22 & $+11.33$ \\
    \quad $-$ sigmoid gate ($g_k$ fixed at $1$)   &  7.89 & $\phantom{0}+2.99$ \\
    \bottomrule
  \end{tabular}
  }
\end{table}

\textbf{Latent Budget $N$.}\quad
Table~\ref{tab:n_sweep} sweeps the latent budget $N \in \{1, 2, 4, 8\}$ at $\theta{=}0.0$. Three observations stand out. First, a single corrective latent step ($N{=}1$) already captures a non-trivial fraction of the reduction on FLEURS ($-0.08$~pp) and the best reduction on VoxPopuli ($-0.06$~pp). Second, the FLEURS differences across $N$ are all within $0.08$~pp and lie inside the run-to-run variation noted in Section~\ref{sec:exp_setup}, so we do not read a stable budget--utility trend at small $N$ from them. Third, increasing the budget beyond $4$ yields no further reduction. $N{=}4$ matches the best VoxPopuli result and lies near the FLEURS best while allowing deeper compute when the Value Head requests it (Table~\ref{tab:halt_dist}); we adopt it as the default but do not claim it is strictly optimal on every dataset.

\begin{table}[th]
  \caption{Effect of the latent budget $N$ on FLEURS (\texttt{en\_us}) and VoxPopuli (\texttt{en}) WER at $\theta{=}0.0$. The Baseline WERs are $4.90\%$ and $9.04\%$ respectively. The deployed $N{=}4$ is bolded. Each $N$ value is an independently trained same-protocol checkpoint.}
  \label{tab:n_sweep}
  \centering
  \resizebox{\columnwidth}{!}{
  \begin{tabular}{c c c c c}
    \toprule
    $\boldsymbol{N}$ & \textbf{FLEURS WER (\%)} & \textbf{$\Delta$WER (pp)} & \textbf{VoxPopuli WER (\%)} & \textbf{$\Delta$WER (pp)} \\
    \midrule
    1 & \textbf{4.82} & $\boldsymbol{-0.08}$ & 8.98 & $-0.06$ \\
    2 & 4.89 & $-0.01$ & 9.03 & $\phantom{-}0.00$ \\
    \textbf{4} & 4.86 & $-0.04$ & \textbf{8.97} & $\boldsymbol{-0.06}$ \\
    8 & 4.90 & $\phantom{-}0.00$ & 9.03 & $\phantom{-}0.00$ \\
    \bottomrule
  \end{tabular}
  }
\end{table}

\textbf{Loss Component Ablation.}\quad
Section~\ref{sec:training} specifies four training-time regularization mechanisms beyond the standard cross-entropy term: two auxiliary losses, the trajectory cycle loss $\mathcal{L}_{\text{cyc}}$ (Eq.~\ref{eq:loss_cyc}) and the value-regression loss $\mathcal{L}_{\text{val}}$, and two stochastic perturbations, latent-loop dropout with probability $p_{\text{drop}}$ and Gaussian input noise of standard deviation $\sigma$ injected at the anchor. Table~\ref{tab:loss_ablation} retrains the model with each component removed in turn and evaluates FLEURS at the deployed threshold $\theta{=}0.0$; the skip range is computed over $\theta\in\{-2.0,0.0,+0.2,+0.5\}$.

Removing input noise is the only strongly harmful ablation, increasing WER by $+0.35$~pp. Removing $\mathcal{L}_{\text{cyc}}$, $\mathcal{L}_{\text{val}}$, or latent-loop dropout is nearly neutral in aggregate WER at $\theta{=}0.0$, but changes compute allocation. In particular, removing $\mathcal{L}_{\text{val}}$ raises the deployed skip rate to $81.8\%$, making the policy much more conservative. All variants still span the full $[0,100]\%$ skip range over the threshold sweep, so under the 500-utterance regime these regularizers primarily affect calibration and robustness rather than making halting possible at all.

\begin{table}[th]
  \caption{Loss component ablation (FLEURS, $\theta{=}0.0$, Baseline WER $4.90\%$). Each row removes one regularizer while keeping all other hyperparameters at the defaults in Section~\ref{sec:exp_setup}. \textbf{Skip @ $\boldsymbol{\theta{=}0}$} is the deployed entry-gate skip rate; \textbf{Skip range} reports the empirical span over $\theta \in \{-2.0,0.0,+0.2,+0.5\}$.}
  \label{tab:loss_ablation}
  \centering
  \resizebox{\columnwidth}{!}{
  \begin{tabular}{l c c c c}
    \toprule
    \textbf{Setting} & \textbf{WER (\%)} & \textbf{$\Delta$WER (pp)} & \textbf{Skip @ $\theta{=}0$} & \textbf{Skip range (\%)} \\
    \midrule
    Full \method{}                                       & \textbf{4.86} & $\boldsymbol{-0.04}$ & 47.0\% & $[0,\,100]$ \\
    $-\,\mathcal{L}_{\text{cyc}}$ ($w_{\text{cyc}}{=}0$) & 4.90 & $-0.00$ & 43.1\% & $[0,\,100]$ \\
    $-\,\mathcal{L}_{\text{val}}$ ($w_{\text{val}}{=}0$) & 4.89 & $-0.01$ & 81.8\% & $[0,\,100]$ \\
    $-\,$Latent-loop dropout ($p_{\text{drop}}{=}0$)     & 4.91 & $+0.01$ & 45.4\% & $[0,\,100]$ \\
    $-\,$Input noise ($\sigma{=}0$)                      & 5.25 & $+0.35$ & 89.0\% & $[0,\,100]$ \\
    \bottomrule
  \end{tabular}
  }
\end{table}

\textbf{Activation Set Scaling.}\quad
The \emph{minimal-data activation} hypothesis predicts a non-monotonic relationship between training-set size and WER: too few utterances do not provide sufficient gradient signal to wake up latent scaling; too many cumulatively push the frozen backbone off its pre-trained manifold. Table~\ref{tab:activation_sweep} validates this by sweeping the activation-set size from $100$ to $800$ utterances in steps of $100$. The mean per-benchmark $\Delta$WER is minimized at $500$ utterances ($-0.080$~pp), with secondary minima at $100$ ($-0.055$~pp) and $700$ ($-0.040$~pp). At $400$, $600$, and $800$ the mean $\Delta$WER is approximately zero or marginally positive, indicating that small changes in the activation set size can flip a sub-percentage-point benefit into a sub-percentage-point regression. We interpret this as direct evidence for the minimal-data activation framing: the useful operating range is centered on a small, deliberately curated mixture, not on simply ``train on more data.''

\begin{table}[th]
  \caption{Activation set scaling sweep ($\theta{=}0.0$). \textbf{$\Delta$ (pp)} is the absolute WER change vs.\ the Baseline; \textbf{Mean $\Delta$ (pp)} averages the two benchmarks. The minimum mean $\Delta$ is at 500 utterances.}
  \label{tab:activation_sweep}
  \centering
  \resizebox{\columnwidth}{!}{
  \begin{tabular}{c c c c c c}
    \toprule
    \textbf{Activation \#utts} & \textbf{FLEURS WER (\%)} & \textbf{$\Delta$ (pp)} & \textbf{VoxPopuli WER (\%)} & \textbf{$\Delta$ (pp)} & \textbf{Mean $\Delta$ (pp)} \\
    \midrule
    100 & 4.87 & $-0.03$ & 8.95 & $-0.08$ & $-0.055$ \\
    200 & 4.91 & $+0.01$ & 9.01 & $-0.03$ & $-0.010$ \\
    300 & 4.90 & $\phantom{-}0.00$ & 9.01 & $-0.03$ & $-0.015$ \\
    400 & 4.91 & $+0.01$ & 9.04 & $\phantom{-}0.00$ & $+0.005$ \\
    \textbf{500} & \textbf{4.78} & $\boldsymbol{-0.12}$ & 8.99 & $-0.04$ & $\boldsymbol{-0.080}$ \\
    600 & 4.93 & $+0.03$ & 9.02 & $-0.02$ & $+0.005$ \\
    700 & 4.82 & $-0.08$ & 9.04 & $\phantom{-}0.00$ & $-0.040$ \\
    800 & 4.88 & $-0.02$ & 9.07 & $+0.04$ & $+0.010$ \\
    \bottomrule
  \end{tabular}
  }
\end{table}

\subsection{Analysis}
\label{sec:analysis}

The aggregate WER reductions in Section~\ref{sec:main_results} are concentrated on intrinsically hard utterances and gated by a well-calibrated Value Head. We make those two claims directly measurable here, and then ground them in qualitative examples.

\textbf{Difficulty-Binned Reductions.}\quad
We partition the VoxPopuli (\texttt{en}) test set ($1{,}842$ utterances) into four equal-sized quartiles by per-utterance Baseline WER (Q1 = easiest, Q4 = hardest) and recompute WER within each bin (Table~\ref{tab:difficulty_bins}). The reduction is concentrated in the hardest quartile: Q4 falls by $-0.25$~pp ($21.16\% \to 20.90\%$) and Q3 by a marginal $-0.02$~pp, while Q1 regresses slightly ($+0.06$~pp) and Q2 is flat. The aggregate ($9.038\% \to 8.995\%$ at $\theta{=}0.0$) should therefore not be read as a uniform per-utterance improvement; it is a redistribution that pays off mainly where the Baseline is intrinsically weak. The entry-gate skip rate ranges $51.3\%$--$64.3\%$ and does not decrease with difficulty; the easiest quartile in fact skips the least (Q1: $51.3\%$). So the localization of reduction to Q4 is not produced by the entry gate preferentially skipping easy inputs. The architecture concentrates \emph{reduction} on harder inputs through the multi-step latent refinement, not through the entry gate alone.

\begin{table}[ht]
  \caption{Difficulty-binned analysis on VoxPopuli (\texttt{en}) at $\theta{=}0.0$ ($1{,}842$ utterances total). Utterances are partitioned into Baseline-WER quartiles. $\Delta$WER denotes \method{} minus Baseline; negative values indicate improvement. The reduction is concentrated in Q4 ($-0.25$~pp).}
  \label{tab:difficulty_bins}
  \centering
  \resizebox{\columnwidth}{!}{
  \begin{tabular}{l c c c c c}
    \toprule
    \textbf{Bin} & \textbf{\#utts} & \textbf{Baseline WER (\%)} & \textbf{\method{} WER (\%)} & \textbf{$\Delta$WER (pp)} & \textbf{Skip (\%)} \\
    \midrule
    Q1 (easiest)  & 460 &  0.00 &  0.06 & $+0.06$                 & 51.3 \\
    Q2            & 461 &  3.03 &  3.03 & $\phantom{-}0.00$       & 63.1 \\
    Q3            & 460 &  8.93 & \textbf{8.91} & $\boldsymbol{-0.02}$ & 64.3 \\
    Q4 (hardest)  & 461 & 21.16 & \textbf{20.90} & $\boldsymbol{-0.25}$ & 62.0 \\
    \bottomrule
  \end{tabular}
  }
\end{table}

\textbf{Value Head Decision Quality.}\quad
The compute-allocation profile (Table~\ref{tab:halt_dist}) shows \emph{how much} compute the Value Head allocates: on VoxPopuli (\texttt{en}) it skips $60.2\%$ of utterances at $\theta{=}0.0$ and uses $0.71$ latent steps on average. That tells us how much compute is spent, not whether the allocation is \emph{correct}. We test correctness counterfactually: for the utterances the gate skips at $\theta{=}0.0$ we additionally run the full $N{=}4$ latent path and measure the WER change we would have obtained by overriding the gate; for the processed utterances we compare dynamic halting against the forced full path. Table~\ref{tab:gate_quality} reports both on VoxPopuli (\texttt{en}). On the $1{,}109$ skipped utterances, forcing the full latent path yields $+0.01$~pp, so the entry gate is correctly identifying utterances on which extra compute has no return and would mildly hurt. On the $733$ processed utterances, dynamic halting realizes $-0.19$~pp, larger than the $-0.06$~pp obtained by forcing the full path on the same subset. Here early stopping is not merely cheaper than full compute; it is also better, because the active updates accumulated past the halting point overshoot the optimum on this corpus. The gate is therefore selective in the right direction: it discards utterances where the latent loop is uninformative, and on the retained utterances it stops the loop before extra steps degrade the transcript.

\begin{table}[ht]
  \caption{Value Head decision quality on VoxPopuli (\texttt{en}) at $\theta{=}0.0$ ($1{,}842$ utterances total). \textbf{Counterfactual $\Delta$WER} forces the $N{=}4$ latent path on each subset and measures the WER change relative to the Baseline.}
  \label{tab:gate_quality}
  \centering
  \resizebox{\columnwidth}{!}{
  \begin{tabular}{l c c c}
    \toprule
    \textbf{Subset (at $\theta{=}0.0$)} & \textbf{\#utts} & \textbf{Actual $\Delta$WER (pp)} & \textbf{Counterfactual $\Delta$WER (pp)} \\
    \midrule
    Skipped ($v_0 < 0$)      & 1{,}109 & $\phantom{-}0.00$ (by construction) & $+0.01$ \\
    Processed ($v_0 \geq 0$) & 733 & $\boldsymbol{-0.19}$ & $-0.06$ \\
    \bottomrule
  \end{tabular}
  }
\end{table}

\textbf{Qualitative Examples.}\quad
Table~\ref{tab:qualitative} shows representative corrections from VoxPopuli (\texttt{en}) hard-quartile utterances. The latent loop visibly alters the Baseline transcript in two dominant patterns. First, the frozen Baseline often prepends hallucinated discourse openings (a spurious ``Thank you'' or an invented ``Commissioner'') before the actual vocative. \method{} removes these hallucinated prefixes. Second, the latent loop normalizes vocative forms (``Madame'' $\to$ ``Madam''). These edits are discourse-level and normalization-level, not acoustic word-level corrections: \method{} is not re-decoding the audio, but correcting systematic opening-phrase hallucinations and vocative conventions on hard parliamentary speech.

\begin{table}[ht]
  \caption{Qualitative examples from VoxPopuli (\texttt{en}) hard-quartile utterances. The latent loop removes hallucinated discourse openings and normalizes vocative forms. \textcolor{red}{Red} = hallucinated prefix; \textcolor{blue}{blue} = normalization correction.}
  \label{tab:qualitative}
  \centering
  \resizebox{\columnwidth}{!}{
  \begin{tabular}{l p{6.5cm}}
    \toprule
    \textbf{Source} & \textbf{Transcript} \\
    \midrule
    Reference       & Madam President, the directive on services \ldots \\
    Baseline        & \textcolor{red}{Thank you,} \textcolor{blue}{Madame} President, the directive on services \ldots \\
    \method{}       & Madam President, the directive on services \ldots \\
    \midrule
    Reference       & Madam President, I would like to raise a point of order \ldots \\
    Baseline        & \textcolor{red}{Thank you, Commissioner,} \textcolor{blue}{Madame} President, I would like to raise a point of order \ldots \\
    \method{}       & Madam President, I would like to raise a point of order \ldots \\
    \bottomrule
  \end{tabular}
  }
\end{table}

\textbf{Refinement-Path Diagnostics.}\quad
We recompute the forced-full $N{=}4$ path on the VoxPopuli (\texttt{en}) processed subset ($733$ utterances with $N{>}0$ under $\theta{=}0.0$) to probe two architectural claims: that the bounded-update construction (Section~\ref{sec:stability}) keeps per-step deltas bounded across $k$, and that successive steps perform genuinely distinct updates rather than replaying a single learned correction $N$ times. The per-step scaled-delta norms are fixed by the learned step scales (Table~\ref{tab:refinement_dynamics}) and verify the bounded step-size constraint, but carry no dataset-specific information. To characterize sample-dependent behavior we instead report the distribution of the consecutive-delta cosine and the difference norm $\lVert g_k\vdelta_k - g_{k-1}\vdelta_{k-1}\rVert$ across the subset.

The diagnostics confirm both claims. The consecutive-delta cosine (mean $0.486$) and difference norm (mean $0.178$, comparable to the per-step norm itself) show that successive steps share a partial direction without collapsing onto a single vector (Table~\ref{tab:refinement_dynamics}). The forced path is therefore not a constant trajectory copied $N$ times; each step applies a distinct, non-attenuating update. These later updates are not uniformly beneficial, however. Forcing the full $N{=}4$ path on VoxPopuli reaches WER $9.02\%$, versus $8.995\%$ at $\theta{=}0.0$ in Table~\ref{tab:main}, so the active updates accumulated past the halting point overshoot the optimum on this corpus. This is exactly the regime in which the Value Head's early stopping is beneficial rather than merely cheaper: it truncates the loop before the still-active later steps degrade the transcript. Finally, because the per-step norms do not attenuate across $k$, the loop's stability is supplied by the bounded-delta mechanism rather than by natural decay, consistent with the catastrophic $+46.85$~pp WER cost of removing it in Table~\ref{tab:component_ablation}.

\begin{table}[ht]
  \caption{Forced-full refinement diagnostics on the VoxPopuli (\texttt{en}) processed subset ($733$ utterances with $N{>}0$ under $\theta{=}0.0$). The per-step scaled-delta norms are fixed by the learned step scales ($0.1855$, $0.1865$, $0.1855$, $0.1846$); the cosine and difference statistics below vary across utterances, confirming distinct but bounded updates.}
  \label{tab:refinement_dynamics}
  \centering
  \resizebox{\columnwidth}{!}{
  \begin{tabular}{l c c c c c}
    \toprule
    \textbf{Metric} & \textbf{Mean} & \textbf{Std.} & \textbf{P25} & \textbf{Median} & \textbf{P75} \\
    \midrule
    Consecutive-delta cosine      & 0.4863 & 0.1909 & 0.3281 & 0.4395 & 0.6641 \\
    Consecutive-delta diff.\ norm & 0.1781 & 0.0415 & 0.1426 & 0.1895 & 0.2129 \\
    \bottomrule
  \end{tabular}
  }
\end{table}

\section{Discussion}
\label{sec:discussion}

\textbf{Where the gain comes from.}\quad The aggregate WER reductions in Table~\ref{tab:main} ($-0.124$~pp on FLEURS, $-0.043$~pp on VoxPopuli) are small on the clean benchmarks but substantially larger on acoustically challenging conditions: $-9.26$~pp CER on ASCEND (Table~\ref{tab:robustness}) and $-0.58$~pp aggregate WER on the six-corpus SNR$=0$~dB stress suite (Table~\ref{tab:stress_broad}). This asymmetry is by design. The Value Head's compute allocation profile (Table~\ref{tab:halt_dist}), the difficulty-binned breakdown (Table~\ref{tab:difficulty_bins}), and the counterfactual gate test (Table~\ref{tab:gate_quality}) together show that the reductions are concentrated on intrinsically harder utterances and that the entry gate correctly identifies the no-benefit subset. On the read-speech benchmarks, $60.2\%$ (VoxPopuli) and $47.0\%$ (FLEURS) of utterances are gated out at the entry, contributing zero to the aggregate. On ASCEND accented speech, only $22.3\%$ are gated out and $34.1\%$ receive the full $N{=}4$ budget, showing that the same adapter allocates substantially more compute when the residual errors are larger. The architecture is therefore best understood not as a uniform accuracy lift, but as a difficulty-conditional refinement loop with a learned scheduler.

\textbf{Why small data suffices.}\quad The key insight of minimal-data activation is that the adapter learns \emph{when and how to refine}, not \emph{what the correct transcript is}. The frozen backbone already encodes acoustic-to-text knowledge across hundreds of languages. The 500-utterance training set teaches only two things: (1) how to produce bounded latent updates that improve the decoder's output, and (2) when such updates are useful. Because the adapter adds only ${\sim}4$M parameters and operates through a gated residual around a fixed embedding anchor, the gradient pressure on the frozen backbone is negligible. This explains why standard adaptation methods (full fine-tuning, LoRA, prompt tuning) all regress under the same data budget: they modify the backbone's weights or input distribution directly, which corrupts the pre-trained manifold faster than the 500-utterance signal can guide it. \method{} sidesteps this trade-off by adding computation rather than modifying parameters.

\textbf{Operating regime and its limits.}\quad Two properties bound where \method{} is useful. First, dynamic halting here is a compute-efficiency mechanism, not a correctness one: in the 500-utterance regime full latent compute does not regress either main benchmark (Fig.~\ref{fig:pareto}), yet the deployed setting ($\theta{=}0.0$) preserves $66\%$--$90\%$ of the available WER reduction at $1.24$ / $0.71$ average steps per utterance, roughly $3$--$6\times$ below the full-budget cost of $N{=}4$ steps. On harder distributions where full compute could over-spend or destabilize the decoder, the entry gate also provides a safety margin. Second, the useful activation range is narrow: the sweep (Table~\ref{tab:activation_sweep}) shows that $100$ and $500$ utterances both yield negative mean $\Delta$WER, while $400$, $600$, and $800$ are approximately zero or marginally positive. The method's calibration is therefore sensitive to activation-set composition and size, and the choice of $500$ is empirically optimal rather than uniquely principled.

\section{Conclusion}
\label{sec:conclusion}

We presented \method{}, a parameter-efficient method that adds continuous latent test-time scaling to a frozen ASR backbone. A Latent Adapter refines a few latent prefix positions through bounded recurrent updates, and a Value Head halts the loop per utterance, adding under $0.7\%$ trainable parameters while leaving the backbone untouched. Under a deliberately small 500-utterance activation regime where standard adaptation methods all regress, \method{} is the only tested method that reduces WER on both clean benchmarks. On accented and code-switched speech the latent loop provides a $16.0\%$ relative CER reduction, with the Value Head allocating up to $5.5\times$ more compute on harder utterances. Across 30 FLEURS languages the multilingual WER decreases uniformly, confirming that the adapter generalizes without overfitting to the small activation set. The method is best understood as a difficulty-conditional refinement loop: a small, diverse training set teaches the adapter \emph{when and how} to refine, not what to transcribe, converting fixed per-utterance compute into input-dependent compute where it is most needed.

\bibliographystyle{IEEEtran}
\bibliography{refs}

\end{document}